\documentclass[preprint,12pt]{elsarticle}

\usepackage{amssymb}
\usepackage{amsmath}

\usepackage{lineno}
\usepackage{color}

\journal{Computer Methods and Programs in Biomedicine}

\begin{document}

\begin{frontmatter}



\title{Incision trajectory tracing for electrosurgical navigation by CNN-based knife contacting frames extraction method} 


\author[inst1]{Yu Chun Wang}
\author[inst2]{Kaixu Chen}
\author[inst3]{Naoto Ienaga}  
\author[inst3]{Yoshihiro Kuroda}

\affiliation[inst1]{organization={Degree Programs in Systems and Information Engineering},
            addressline={University of Tsukuba}, 
            city={Tsukuba},
            postcode={305-8573}, 
            state={Ibaraki},
            country={Japan}}

\affiliation[inst2]{organization={Center for Computational Sciences},
            addressline={University of Tsukuba}, 
            city={Tsukuba},
            postcode={305-8577}, 
            state={Ibaraki},
            country={Japan}}

\affiliation[inst3]{organization={Institute of Systems and Information Engineering},
            addressline={University of Tsukuba}, 
            city={Tsukuba},
            postcode={305-8573}, 
            state={Ibaraki},
            country={Japan}}
            
\begin{abstract}
Background and Objective: Image-guided surgical navigation has been actively studied because of its advantage of identifying subsurface targets and critical structures, whereas it requires incision trajectories to update the preoperative three-dimensional model dynamically during the surgery. The novelty of this study is the thermal feature distinguishment of whether the electric tools contacting the tissue by Convolutional Neural Network (CNN), and the extraction of the knife contacting frames, to form incision trajectories which can meet with the requirement during the surgery. 
\\Methods: This study firstly verified that CNN can classify the thermal images of electric knife and ultrasonic cutter operations separately, and can raise the accuracy of the incision trajectories derived from the connection of the thermal intensity centroid of the frames predicted by CNN as contacting. 
\\Results: Our results obtained by employing the electric knife not only reveal a remarkably high accuracy 97.2~\% in CNN’s identification, but also can achieve an error reduction as high as more than 2.5 times of the incision trajectory prediction as compared to those proceeded in the conventional method. Besides electric knife, the results obtained by employing another electric tool, ultrasonic cutter, reveal a high accuracy up to 93.7~\%. 
\\Conclusion: In this study, we ensured the possibility of CNN in distinguishing electric tools contacting with the tissue, and confirmed that the proposed method has not only overcome the problem of missing trajectories which usually occurs in the convolutional long-short term memory method but also achieved a remarkable improvement of the accuracy with less limitation. 
\end{abstract}



\begin{keyword}{
CNN \sep Surgical navigation \sep Thermal image \sep Electrosurgery \sep Trajectory \sep Machine learning
}



\end{keyword}

\end{frontmatter}



\section{Introduction}
Recently, image-guided surgery (IGS), specifically augmented reality (AR), virtual reality (VR), and mixed reality (MR) technologies navigation has been actively studied~\cite{Z. Asadi} in order to overcome the incomplete vision during the complex surgical procedure which inhibits the understanding on the internal structure of an organ. Although indocyanine green (ICG) fluorescence imaging has been regarded as a common navigation method, it merely focuses on the surface tumor identification and the segmentation. On the other hand, MR IGS has special advantage of identifying the key anatomical structure and guiding the resections by displaying incision trajectories and margins to avoid vascular injuries and neurological deficits~\cite{S. Bernhardt}, it also has revealed helpful improvements such as shorter operative time and less blood loss~\cite{N. Zeng}. 

For adapting a dynamic change of the surgical scene, it is necessary to acquire the exact location of the incision and update the preoperative three-dimensional (3D) model dynamically for the IGS~\cite{N. Haouchine}. A conventional approach was reported to use feature points on the organ surface to estimate the incision~\cite{C. J. Paulus}. However, feature points could not always be detected properly due to specular reflection and blood adhesion. Kim et al.~\cite{J. W. Kim} proposed a navigation method for surgical tools based on training a CNN to track the tool-tip. However, the network training relies on images of the tool and its shadow convergence. In other words, a clear tool shadow is required throughout the entire navigation process, which limits the applicability of this method.

\begin{figure}
        \centering        \includegraphics[width=0.5\textwidth]{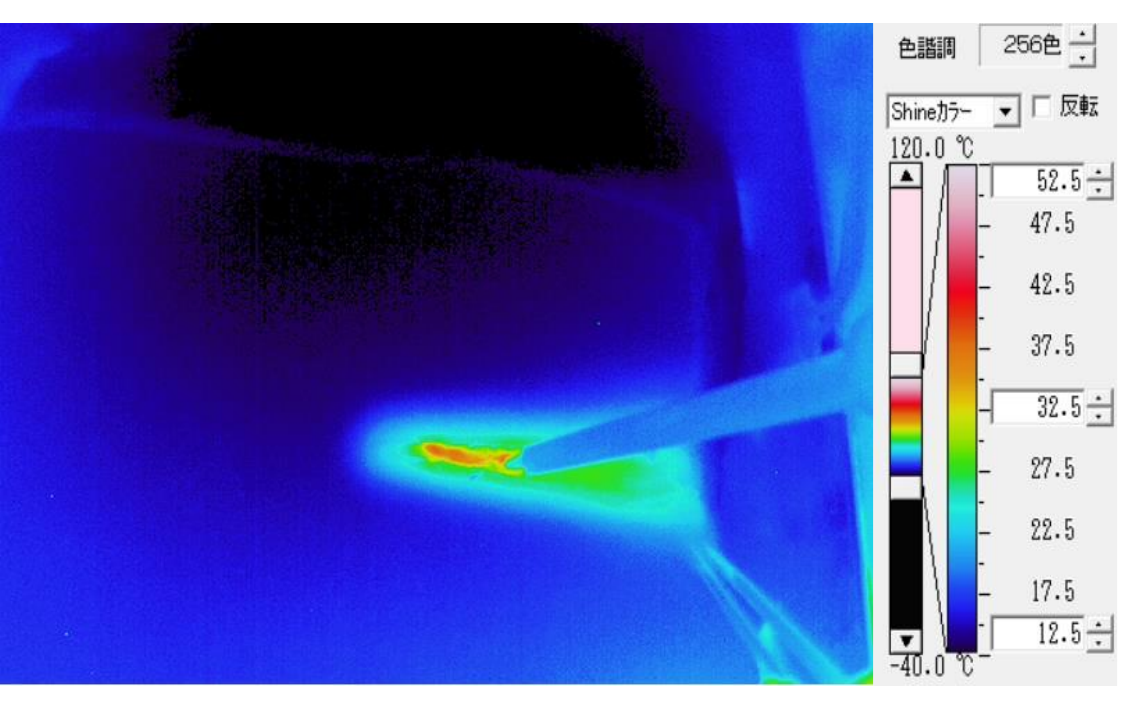}
        \caption{Cavitron Ultrasonic Surgical Aspirator (CUSA) left thermal diffusion on the tissue acquired by thermal camera. The temperature on the graph is measured in degrees Celsius.}
        \label{Fig. 1}
    \end{figure}

For the surgical operation conducted with the electric device such as electric knife or Cavitron Ultrasonic Surgical Aspirator (CUSA)~\cite{J. M. Little}, it was found that the high temperature of the electric device raised tissue temperature rapidly as shown in Figure~\ref{Fig. 1}. Mizunuma et al.~\cite{Y. Mizunuma} proposed a Convolutional Long-Short Term Memory (ConvLSTM) based incision trajectory estimation method which can track the post contact diffusion caused by the electric knife from thermal image sequence to avoid the false detection on tracing the electric knife that has no contact with the tissue. 
Nevertheless, several unsolved points remained in the ConvLSTM method. For example, ConvLSTM network was suggested to set the prediction for the trajectory multiple frames earlier which may cause a five second delay~\cite{Y. Mizunuma}. Moreover, the model focused on post-contact diffusion images only, and thus requires the removal of the frames containing electric knife to maintain the accuracy, but the requirement of the electric knife disengagement yields a non-continuous navigation. Such kind of weak points can be mitigated potentially if more thermal features, such as the electric knife contacting feature, can be recognized and used efficiently.

This study aims to create a novel frame extracting method combining both the approach of training a Convolution Neural Network (CNN) to classify the thermal images of the electric knife and ultrasonic cutter’s cutting status (i.e. identifying whether it contacts the tissue or not) and the approach of acquiring the highest temperature point in each frame to overcome the weak points such as the delay that occurs in the ConvLSTM research so as to provide a higher accuracy. 
The contributions of our research are mainly in two aspects: 

    \begin{itemize}
        \item We firstly proposed a method which can verify the possibility of CNN in distinguishing the real contact status between electric knife and/or ultrasonic cutter and the tissue.
        \item The advantage of our method generating the incision trajectories can provide a higher accuracy requiring neither the succeeding frames nor the removal of electric knife containing images in advance as those are required in the ConvLSTM method.

    \end{itemize}
    
\section{Related research}

IGS showed its advantages in reducing workload by saving the surgeon from mentally matching information of different sources~\cite{M. N. Theodoraki} and exemption from intraoperative complications~\cite{D. M. Dalgorf}. In addition, IGS can also be a useful tool to train novice surgeons~\cite{I. L. Schmale}. Therefore, various surgical domains have been studied with MR IGS such as orthopedic~\cite{A. J. Butler,F. R. Bhatt}, neurosurgery~\cite{B. Carl,M. H. Bopp}, endoscopic~\cite{J. Zeiger,M. Linxweiler},  hepatectomy~\cite{Y. Saito,I. M. Sauer} and craniomaxillofacial~\cite{M. Zhu}. 

MR used in IGS usually contains the following steps. First, acquiring patient-specific data from computed tomography (CT)~\cite{R. Tang}, magnetic resonance imaging (MRI)~\cite{N. Bourdel}, positron emission tomography (PET)~\cite{J. Liu}, or creating a preoperative model by combining the above methods~\cite{R. Galati}. Second, visualizing the target structure such as the tumor, the relevant anatomy around the target structures, and the risk structures such as blood vessels. Next, surface rendering virtual data such as surface rendering or volume rendering~\cite{D. Bartz}, and overlaying the imaging data for an intuitive understanding of the relationship to clarify how various structures are related in 3D space. Last, present the information to the surgeon via display such as monitor~\cite{Y. Goto}, head-mounted display (HMD)~\cite{C. Scherl} or microscope~\cite{Z. G. Schwam}. It should be noted that updating the 3D model during surgery is required in the last step. 

Most of the researches focused on RGB images to acquire the intraoperative changes, but false detection via RGB images can be caused by blood adhesion, light reflection, and insufficient brightness~\cite{T. A. Alshirbaji}. These defects can be mitigated by an alternative solution using thermography to provide information different from RGB. However, most of the machine learning studies on thermal medical images are limited to disease identifications such as the detections of breast cancer~\cite{J. Zuluaga-gomez,S. J. Mambou,V. R. Allugunti}, facial paralysis~\cite{X. Liu} and carotid artery stenosis~\cite{A. Saxena}. Mizunuma et al.~\cite{Y. Mizunuma} thereby stated a ConvLSTM based estimation method for electric knife by focusing on the thermal diffusion of the incision trajectory to improve the machine learning on thermal trajectory prediction in surgical applications. 

Although Brahmbhatt, et al.~\cite{S. Brahmbhatt} has proposed the human grasp contact dataset via thermal image, which verifies the heat conduction from human hands to the object via the contact between hands and the object, yet the thermal conduction and the thermal contact from electric medical equipment to the tissue by the CNN recognition have not been reported so far. 

The appearance of ResNet~\cite{K. He} solved the problem of degradation in CNN. This allows the network to learn more features and reach a higher accuracy. Since then, the fields of computer vision and image analysis have made significant progress. Then, EfficientNet~\cite{M. Tan}, an extension of ResNet, was developed, which can balance the network width, depth, and resolution in a simple but effective way. In this study, we further develop a method by training the EfficientNet with thermal images to generate a novel effect on recognizing the feature of electric cutting devices contacting with the tissue in a higher accuracy.

\section{Methods and materials}
\begin{figure}
        \centering        \includegraphics[width=0.95\textwidth]{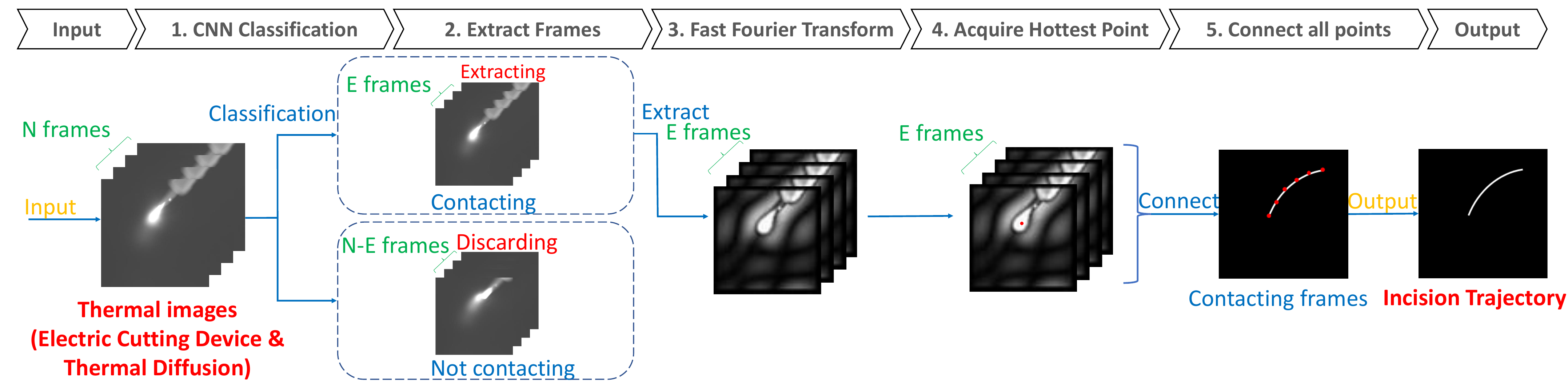}
        \caption{Proposed Method: 1. After input the thermal images, each frame was classified to whether the electric cutting device was “contacting" or “not contacting" the tissue by the CNN. 2.The frames predicted as “contacting" the tissue were extracted. 3.The extracted frames were processed by Fast Fourier Transform (FFT) to enhance the contact area apart from the other part of the electric cutting device. 4. The hottest spots, the tip of the electric cutting device, on each of the extracted frame acquired out by an intensity centroid method. 5. The hottest spots from each frame were connected to form and output as a trajectory.}
        \label{Fig. 2}
    \end{figure}

In order to investigate whether CNN can distinguish the feature of electric cutting tool contacting with the tissue, as well as generate a decent trajectory, we proposed the following thermal diffusion recognition method by CNN shown in Figure~\ref{Fig. 2}. 

\begin{figure}
        \centering        \includegraphics[width=0.7\textwidth]{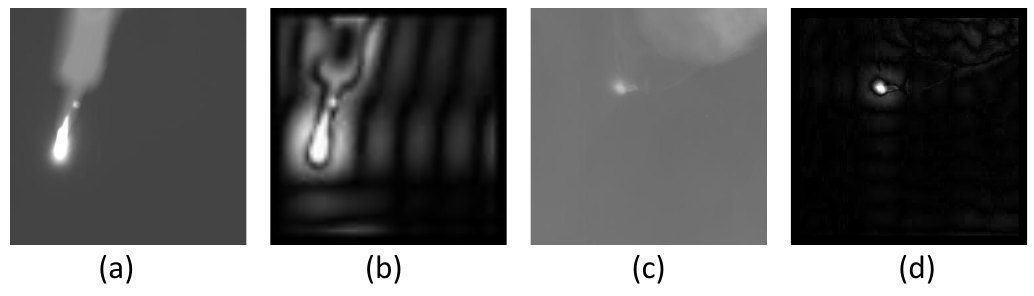}
        \caption{examples of FFT post-processing: (a) The example of electric knife’s frame labeled as “contacting" before FFT post-processing. (b) The example of the electric knife’s frame after FFT post processing.  (c) The example of ultrasonic cutter’s frame labeled as “contacting" before FFT post-processing. (d) The example of the ultrasonic cutter’s frame after FFT post processing. }
        \label{Fig. 3}
    \end{figure}

\begin{figure}
        \centering        \includegraphics[width=0.3\textwidth]{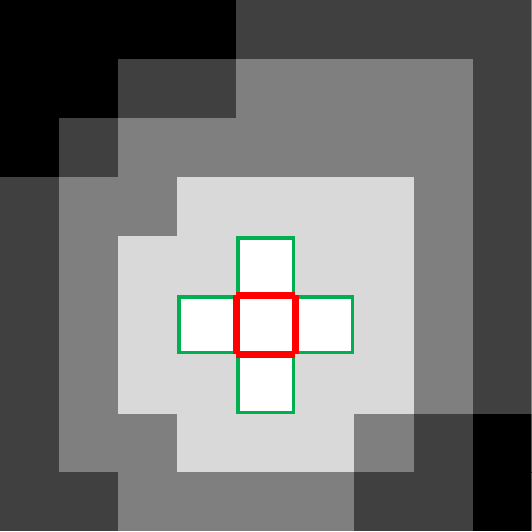}
        \caption{Examples of intensity centroid method: The white pixels with the green boundary shows the hottest pixels of the thermal image, and the pixel with the red boundary shows the hottest point calculate by the intensity centroid method by setting the threshold at the highest temperature at the frame.}
        \label{Fig. 4}
    \end{figure}

Different from the conventional method, which needs to remove the electric knife images in advance, we not only used the images containing thermal diffusion but also used the thermal images containing electric knife. The detailed procedure includes several steps. Firstly, after the input of thermal images, each frame was classified to whether the electric knife was “contacting” with the tissue or “not contacting" with the tissue by CNN. Secondly, the frames predicted as “contacting" with the tissue were extracted because the incision trajectories must be contacted by the electric knife during the operation. Thirdly, the extracted frames were processed by Fast Fourier Transform (FFT) to enhance the contact area apart from the other area. The example of the FFT post-processing is indicated in Figure~\ref{Fig. 3}, including (a) the example of electric knife’s frame labeled as “contacting" before FFT post-processing, (b) the example of the electric knife’s frame after FFT post processing, (c) the example of ultrasonic cutter’s frame labeled as “contacting" before FFT post-processing, and (d) the example of the ultrasonic cutter’s frame after FFT post processing. Then, the hottest spots, i.e. the tip of the electric knife, on each of the extracted frames were acquired by an intensity centroid method. The example of the intensity centroid is disclosed in Figure~\ref{Fig. 4} in which the white pixels with a green boundary represent the hottest pixels of the thermal image, and the pixel with a red boundary represents the hottest point calculated by the intensity centroid method by setting the threshold at the highest temperature at the frame. Finally, the hottest spots from each frame were connected to form and output as a trajectory.    

To acquire the highest temperature point at each frame, we adopted the intensity centroid method to the pixels which are within the range of one percent of the highest temperature in each frame and then output the intensity centroid coordinate at each extracted frame as follows:
\begin{eqnarray}
(\widetilde{x_k},\widetilde{y_k})=\left ( \left[\frac{\sum\limits_{y=1}^{h}\sum\limits_{x=1}^{w}I(x,y)x}{\sum\limits_{y=1}^{h}\sum\limits_{x=1}^{w}I(x,y)}\right ],\left[\frac{\sum\limits_{y=1}^{h}\sum\limits_{x=1}^{w}I(x,y)y}{\sum\limits_{y=1}^{h}\sum\limits_{x=1}^{w}I(x,y)}\right ] \right )
\end{eqnarray}
where $\widetilde{x_k}$ and $\widetilde{y_k}$ are the $x$, $y$ coordinates of the intensity centroid over 60°C at frame k, respectively. $I(x,y)$ is the intensity at the pixel $(x,y)$, and $h$ and $w$ are the height and width of the image, respectively. The incision trajectory is formed by connecting the calculated intensity at each frame.

Since the intensity centroid method cannot distinguish whether or not the electric tools are contacting with the tissue, the trained CNN model for extracting the “contacting" frame was conducted beforehand. By combining CNN classification with FFT post-processing and intensity centroid method, it is expectable to output the incision trajectory with the frames in which the electric knife is contacting with the tissue.

\subsection{Electric tools}
In this study, two electric tools for tissue cutting were employed in two experiments separately for creating the following dataset described in section 3.2. These tools comprise an electric knife (DEL1, Bovie Medical Corporation) for cutting the processed ham commercially available, and an ultrasonic cutter (EchoTech, ZO-80II) for cutting the chicken thigh commercially available.

\subsection{Dataset}
Two datasets obtained from the images collected by the electric knife’s trajectories and the ultrasonic cutter’s trajectories separately, are used to input to the proposed method for evaluation.

\subsubsection{Electric Knife’s trajectories dataset}
\begin{figure}
        \centering        \includegraphics[width=0.5\textwidth]{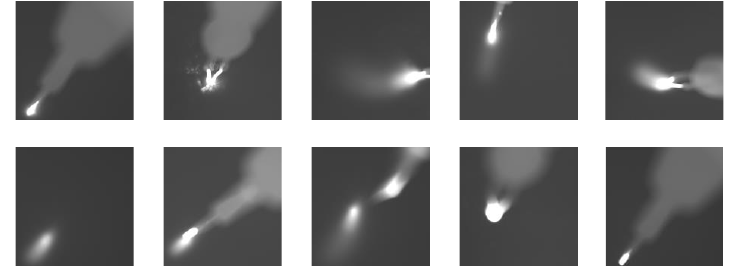}
        \caption{Examples of frame labeled by employing the electric knife: (Top row) The examples of frames labeled as “contacting" and (Bottom row) the examples of frames labeled as “not contacting.}
        \label{Fig. 5}
    \end{figure}

The electric knife was employed to compare the results obtained from both the ConvLSTM method on thermal diffusion images only and the intensity centroid method without the CNN applied. Total 50 sequences of the trajectories were prepared, which comprise 2,141 frames and 3,791 frames labeled as “contacting” and “not contacting,” respectively as shown in Table~\ref{table:Electric Knife’s trajectories dataset}. The path of the incision trajectory was determined in advance, and the ground truth of the incision trajectory was obtained by temporarily placing the undry painted cotton thread on the tissue between which the temperature is different (e.g. the processed ham commercially available is at 4°C and the undry painted cotton thread at 26°C) and then recording with a thermal camera (InfRec R450); the camera lens was placed 11 cm above the top of the tissue surface. The ground truth of the incision trajectory was obtained by manually annotating the thermal trajectories recorded in the thermal image as a landmark. After the cotton thread was removed and the surface of the tissue had reached a uniform temperature, the electric knife was employed to cut the tissue for 3 to 5 times partially during one cutting sequence along with the trajectory of the painted surface derived from the placement of the undry painted cotton thread. In between every partial cut, the electric knife was removed from the object for a short moment. The thermal images were recorded at a frame rate of 2 FPS using a thermal camera with a resolution of $480\times360$ pixels. The length per one pixel was about 0.1~mm. The thermal image sequence was then normalized to [0, 255] in the range from 5 ° C to 60~°C. The work of labeling was conducted manually. The examples of “contacting" frames and “not contacting" frames are shown in Figure~\ref{Fig. 5}. It is possible that mislabels may occur at the moment that the electric knife just contacts or just disengages the object. Even so, the timing and the electric knife movement are considered very short. Consequently, the mislabel will obviously not affect the prediction trajectories theoretically. However, due to the above uncertainty, the final evaluation will be conducted by comparing the prediction trajectory with the predetermined trajectory instead of the accuracy of the CNN classification. 

       \begin{table}[h]
         \caption{Electric Knife’s trajectories dataset}
         \label{table:Electric Knife’s trajectories dataset}
        \centering
          \begin{tabular}{   c    c     c     c    }
           \hline
           \textbf{} & \textbf{Labeled as contacting}& \textbf{Labeled as not contacting}\\
           \hline 
           \textbf{Frames} & 2,141 (36.1\%)&3,791 (63.9\%) \\   \hline
          \end{tabular}
        \end{table}

\subsubsection{Ultrasonic cutter’s trajectories dataset}
\begin{figure}
        \centering        \includegraphics[width=0.5\textwidth]{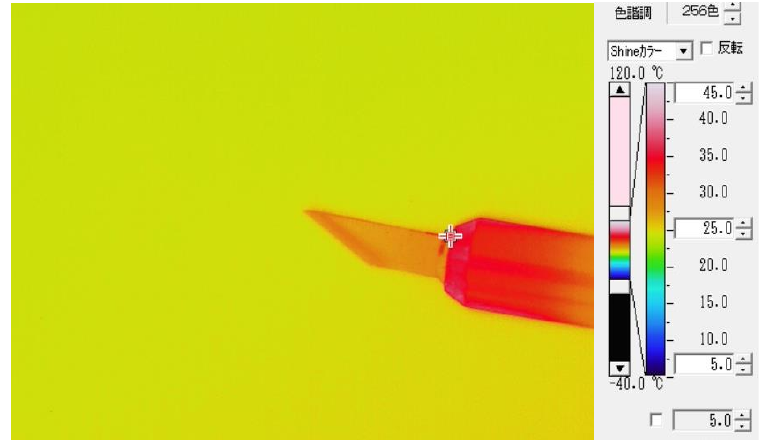}
        \caption{The hottest point of the ultrasonic cutter is not at the tip of the knife}
        \label{Fig. 6}
    \end{figure}

\begin{figure}
        \centering        \includegraphics[width=0.5\textwidth]{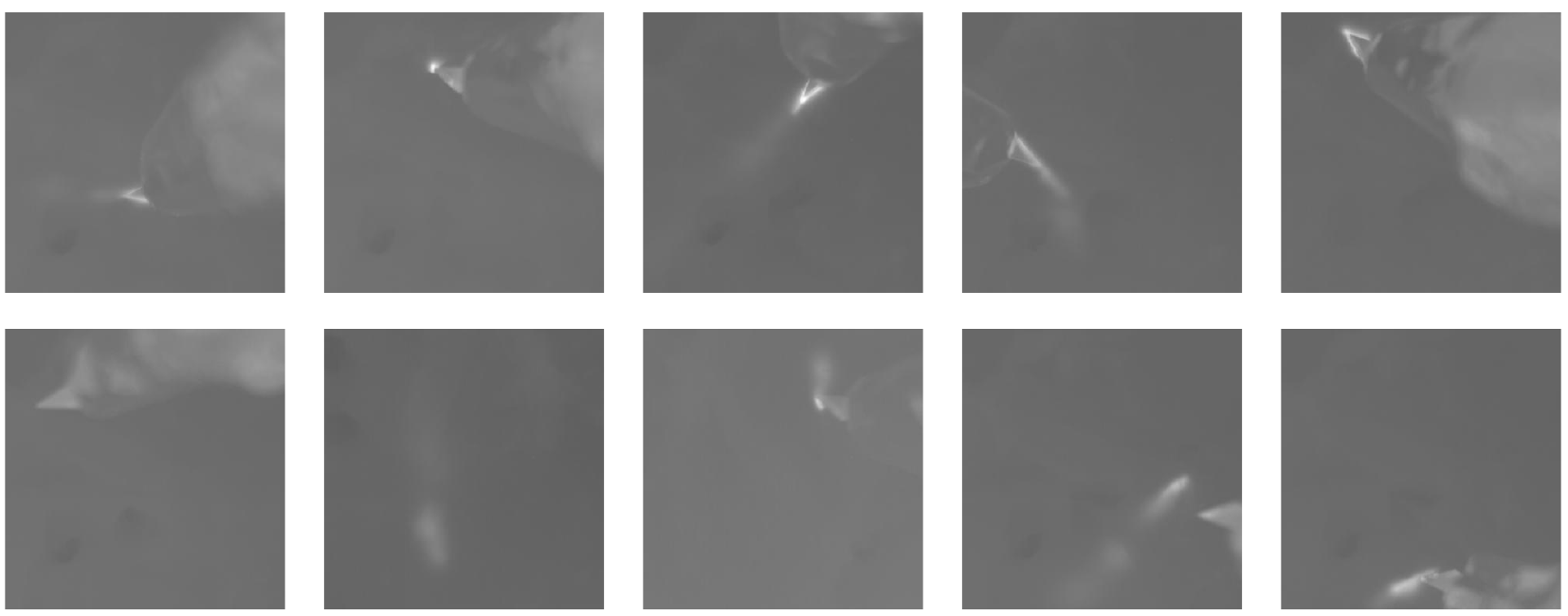}
        \caption{Examples of frame labeled by employing ultrasonic cutter: (Top row)The examples of frames labeled as “contacting" and (Bottom row) the examples of frames labeled as “not contacting.}
        \label{Fig. 7}
    \end{figure}

Since there are other kinds of electrosurgical equipment in the surgical scene, such as the CUSA medical ultrasonic knife, it is important to evaluate whether or not the other kind of thermal diffusion characteristic can be distinguished by CNN. We thereby tried the further evaluation on the other thermal phenomenon by employing the ultrasonic cutter purchased from the commercial market. We employed the ultrasonic cutter at the "high" power setting under thermal camera recording and obtained the result shown in Figure~\ref{Fig. 6}, in which the hottest point of the ultrasonic cutter is not found at the tip of the knife. In contrast, it is found that the contacting point by the CUSA medical ultrasonic knife becomes the hottest point on the tissue. Since the primary purpose of this experiment is to verify the thermal diffusion recognition by CNN, we modified the ultrasonic cutter by covering it with aluminum tape (Nitoms Aluminum Foil Adhesive Tape) which can reduce the thermal radiation. Total 45 sequences of the trajectories were prepared including 2,967 frames and 3,177 frames labeled as “contacting” and “not contacting”, respectively as shown in Table~\ref{table:Ultrasonic cutter’s trajectories dataset}. The path of the incision trajectory was determined in advance by placing the cotton thread at different temperatures in the tissue. To adopt a tissue more similar to human tissue, we used chicken thigh meat instead of commercially available processed ham and conducted the procedures as the same as those in the preparation of the electric knife dataset trajectories except that the chicken thigh meat is at 26°C and the undry painted cotton thread at 4°C. Afterwards, thermal image sequence was normalized to [0, 255] in the range from 5 ° C to 45~°C. The labeling work was performed manually. The examples of “contacting" frames and “not contacting" frames are shown in Figure~\ref{Fig. 7}. As mentioned previously, it is possible that mislabels may occur at the moment that the electric knife just contacts or just disengages the object. Even so, the timing and the ultrasonic cutter movement are considered very short. Consequently, the mislabel will not obviously affect the prediction trajectories theoretically. However, because of the above uncertainty, the final evaluation will be conducted by comparing the prediction trajectory with the predetermined trajectory instead of the accuracy of the CNN classification.

       \begin{table}[h]
         \caption{Ultrasonic cutter’s trajectories dataset}
         \label{table:Ultrasonic cutter’s trajectories dataset}
        \centering
          \begin{tabular}{   c    c     c     c    }
           \hline
           \textbf{} & \textbf{Labeled as contacting}& \textbf{Labeled as not contacting}\\
           \hline 
           \textbf{Frames} & 2,967 (48.3\%)&3,177 (51.7\%) \\   \hline
          \end{tabular}
        \end{table}

\subsection{CNN Training Environment}
   A binary classification was performed for the prediction of “contacting" and “not contacting" by CNN. The CNN used in this study was EfficientNet b4. The input thermal images were center cropped to $360\times360$ pixels. We adopted Adam~\cite{D. P. Kingma} with a learning rate of $1.0\times10^{-3}$  and reduced the learning rate if the accuracy does not improve for 3 consecutive epochs. The adopting batch size was 16. The training was conducted by the pre-trained model on the ImageNet. The implemented system consists of PC with NVIDIA GeForce GTX 1080Ti 12GB, and PyTorch 1.10.2 framework.
   
\subsection{Evaluation Method}
\begin{figure}
        \centering        \includegraphics[width=0.5\textwidth]{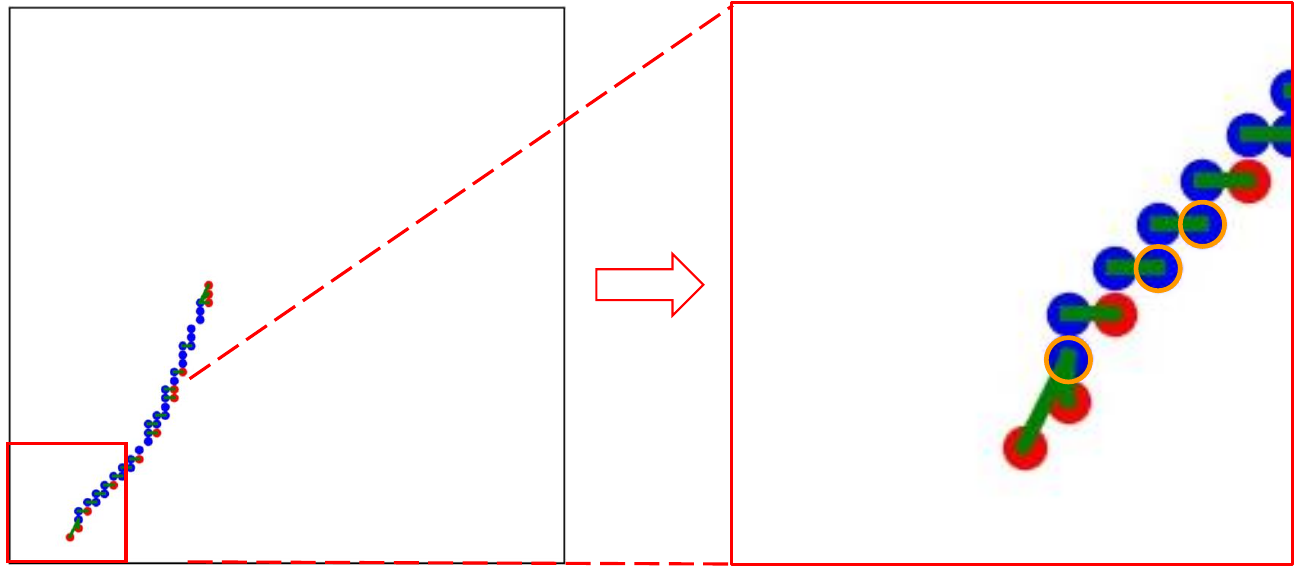}
        \caption{The example of DTW. The red pixels represent the ground truth trajectory, the blue pixels represent the prediction trajectory the orange circles represent the overlap of the pre-determined trajectory in red pixels and the prediction trajectory in blue pixels, and the green lines are the minimum distances between each blue and red point.}
        \label{Fig. 8}
    \end{figure}

In the above experiment, three different prediction methods were performed to the same trajectory dataset, including the ConvLSTM method on thermal diffusion images only, the intensity centroid method without CNN applied, and the proposed method by CNN. 
To evaluate the output trajectories objectively, Dynamic Time Warping (DTW)~\cite{E. J. Keogh} as shown in Figure~\ref{Fig. 8} was utilized to calculate the distance between the prediction and the pre-determined trajectory, which was then divided by the total amount of the path distance to obtain an average distance. The red pixels represent the pre-determined trajectory, the blue pixels represent the prediction trajectory, the orange circles represent the overlap of the pre-determined trajectory in red pixels and the prediction trajectory in blue pixels, and the green lines are the minimum distances between each blue and red point. 

\section{Results}

\subsection{Results of electric knife’s trajectories dataset}
\begin{figure}
        \centering        \includegraphics[width=0.5\textwidth]{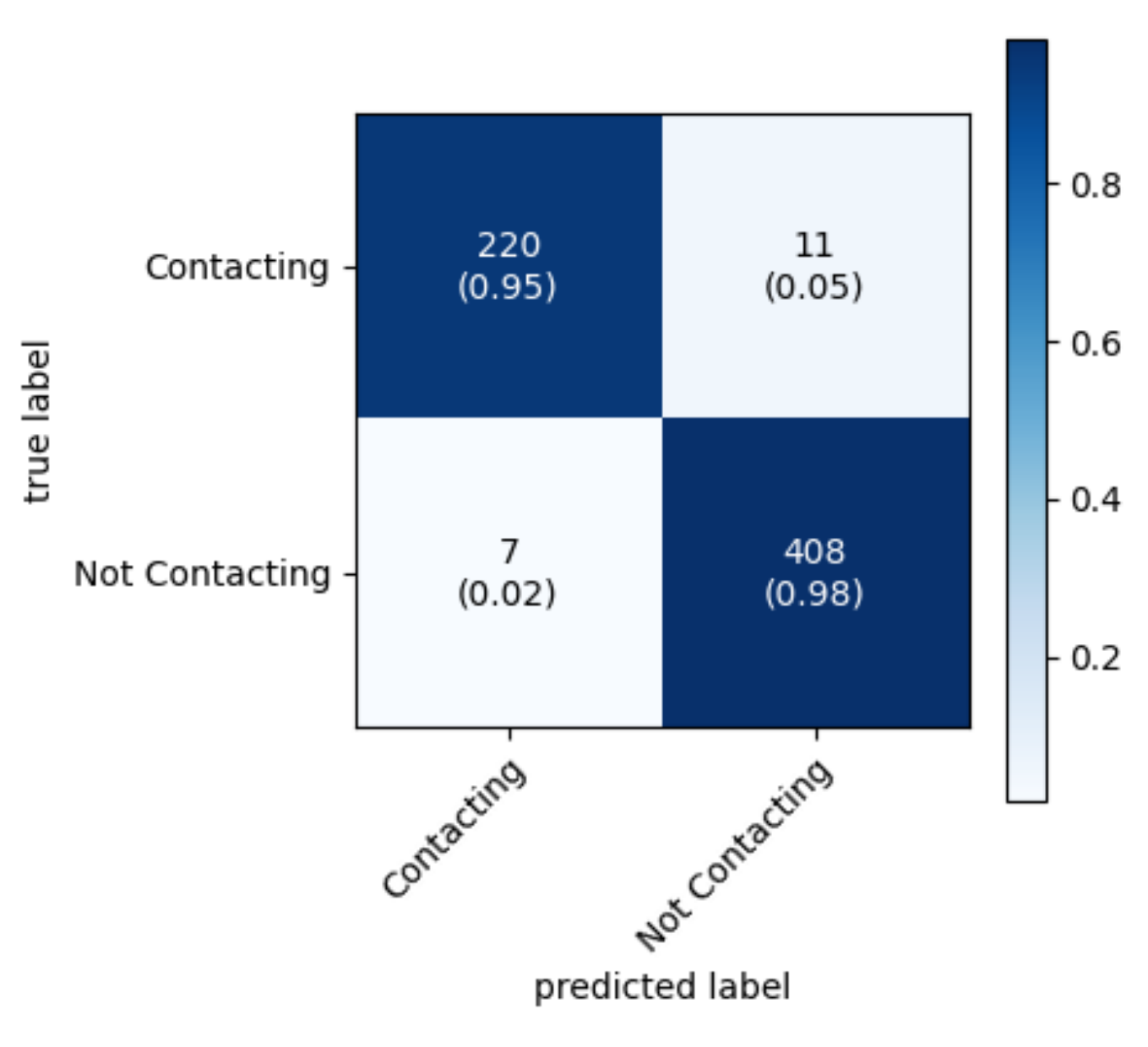}
        \caption{The confusion matrix of the CNN prediction result by employing electric knife, where the top-left number represents the true positives (TP), with the value in parentheses indicating the true positive ratio. The bottom-left number shows the false positives (FP). The top-right number displays the false negatives (FN). Finally, the bottom-right number denotes the true negatives (TN).}
        \label{Fig. 9}
    \end{figure}

\begin{figure}
        \centering        \includegraphics[width=0.5\textwidth]{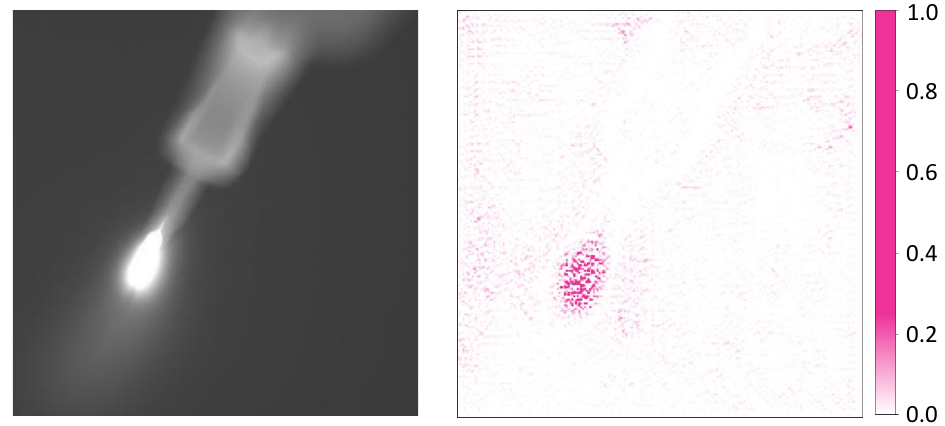}
        \caption{Output from gradient-based attribution by employing electric knife: An example of (Left) the input thermal image, (Right) the output from gradient-based attribution shows the model focusing on the feature at the area of the tip of electric knife and the diffusion area. The density of 0 to 1 represent the positive relevance of the area.}
        \label{Fig. 10}
    \end{figure}

The result of CNN prediction shows an accuracy up to 97.2~\% and F1 score 97~\%, including 628 correct prediction images and 18 incorrect prediction images. The confusion matrix is shown in Figure~\ref{Fig. 9}, where the top-left number represents the true positives (TP), with the value in parentheses indicating the true positive ratio. The bottom-left number shows the false positives (FP). The top-right number displays the false negatives (FN). Finally, the bottom-right number denotes the true negatives (TN).
In order to further interpret and visualize the training results, gradient-based attribution~\cite{M. Sundararajan} was performed on the input thermal images. The output example in Figure~\ref{Fig. 10} shows the model focusing on the feature at the area of the tip of electric knife and the thermal diffusion area which is the same as what we anticipated.

\begin{figure}
        \centering        \includegraphics[width=0.5\textwidth]{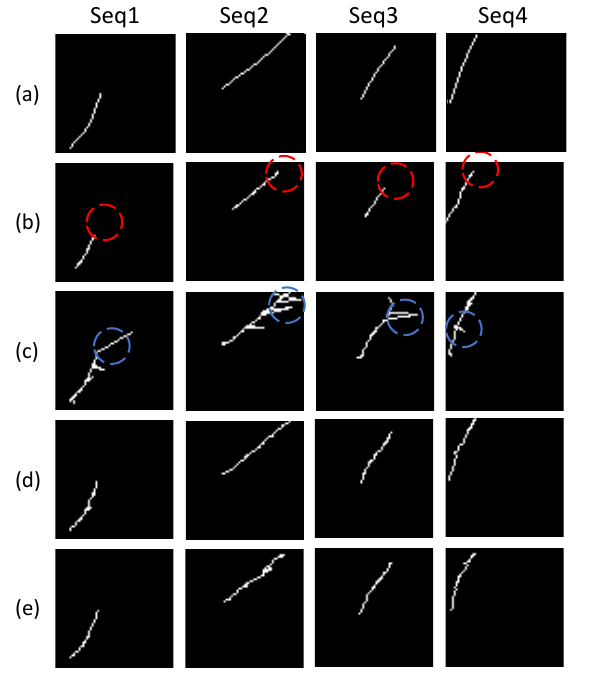}
        \caption{Examples of output by employing electric knife: Examples of output from (a) Ground truth. (b) The output obtained from ConvLSTM method. The red circles point out the missing trajectories compared to ground truth. (c)The output of intensity centroid of the electric knife without CNN applied. The blue circles point out the false detection trajectories compared to ground truth. (d) The proposed method without the step of FFT post-processing. (e) The proposed method with the step of FFT post-processing}
        \label{Fig. 11}
    \end{figure}

In order to compare our proposed method with ConvLSTM method using the resolution of $64\times64$ pixels by~\cite{Y. Mizunuma}, all of our output trajectories were converted into the same resolution. The results regarding the incision trajectory output from ConvLSTM, the intensity centroid of electric knife without CNN applied, and the proposed method are shown in Figure~\ref{Fig. 11}. Table~\ref{table:1} shows the comparison data of DTW. The DTW of ConvLSTM method was average 1.46 pixels with the standard deviation of 0.82 pixels and the obtained data by the intensity centroid methods without CNN were 2.19 pixels with 1.34 pixels deviation, whereas the obtained data by our proposed method without and with the FFT post-processing were 0.55 pixels with 0.13 pixels deviation, and 0.75 pixels with 0.22 pixels deviation, respectively.

       \begin{table}[h]
         \caption{Electric knife’s Incision Trajectory’s DTW. Values are mean ± std.}
         \label{table:1}
        \centering
          \begin{tabular}{   c    c     c     c    }
           \hline
           \textbf{Prediciton Method}& \textbf{Ave DTW [px]} \\
           \hline \hline
           ConvLSTM &1.46±0.82 \\ 
             Intensity Centroid & 2.19±1.34 \\ 
            \textbf{Proposed w/o FFT}  &  \textbf{0.55±0.13} \\
            \textbf{Proposed w/ FFT}  &  \textbf{0.75±0.22}\\ \hline
          \end{tabular}
        \end{table}

\subsection{Results of ultrasonic cutter’s trajectories dataset}
\begin{figure}
        \centering        \includegraphics[width=0.5\textwidth]{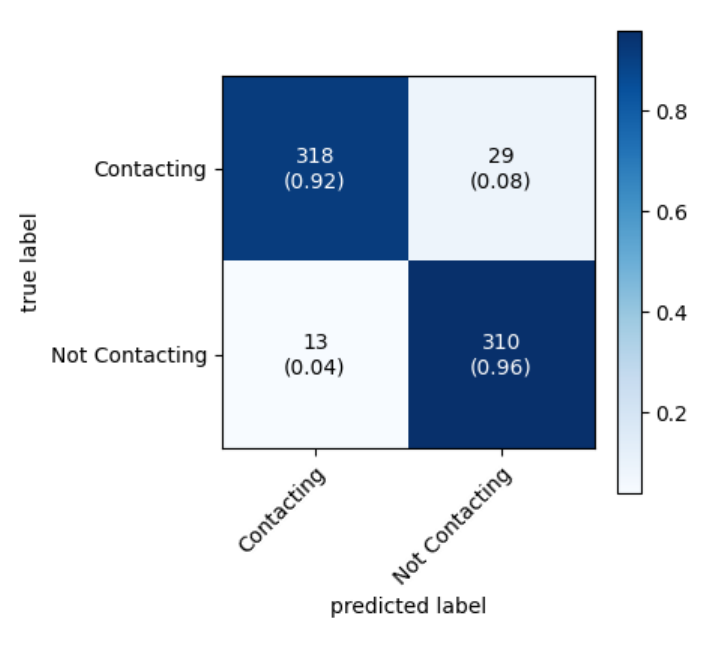}
        \caption{The confusion matrix of the CNN prediction result by employing ultrasonic cutter}
        \label{Fig. 12}
    \end{figure}

The CNN prediction result shows an accuracy of 93.7~\% and F1 score 93.7~\%, including 628 correct prediction images and 42 incorrect prediction images. The confusion matrix result is shown in Figure~\ref{Fig. 12}. 

\begin{figure}
        \centering        \includegraphics[width=0.5\textwidth]{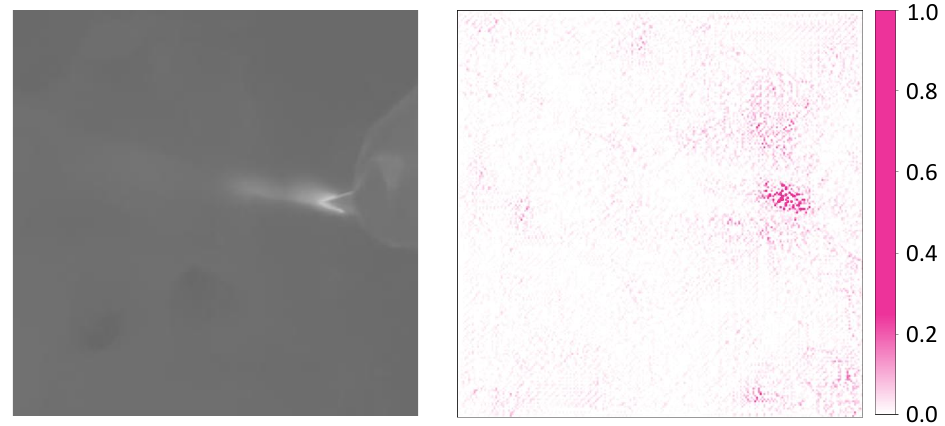}
        \caption{Output from gradient-based attribution by employing ultrasonic cutter: An example of (Left) the input thermal image, (Right) the output from gradient-based attribution shows the model focusing on the feature at the area of the tip of ultrasonic cutter and the diffusion area. The density of 0 to 1 represent the positive relevance of the area.}
        \label{Fig. 13}
    \end{figure}

\begin{figure}
        \centering        \includegraphics[width=0.5\textwidth]{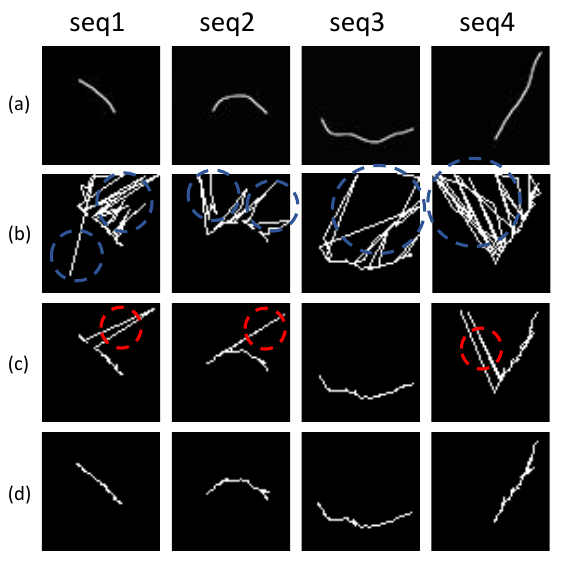}
        \caption{Examples of output by employing electric knife: (a) The ground truth trajectories (b)The output of intensity centroid of the ultrasonic cutter without CNN applied. The blue circles point out the false detection trajectories compared to ground truth. (c)The proposed method’s prediction trajectories without FFT post-processing. The red circles point out the false connecting point due to the higher temperature on the ultrasonic cutter body. (d)The proposed method’s prediction trajectories with FFT post-processing }
        \label{Fig. 14}
    \end{figure}

In order to further interpret and visualize the trained model, gradient-based attribution as performed on the input thermal images. The output example in Figure~\ref{Fig. 13} shows the model focusing on the feature at the area of the tip of ultrasonic cutter and the thermal diffusion area which is similar to that of electric knife. Figure~\ref{Fig. 14} reveals the results regarding the incision trajectory output from the intensity centroid of ultrasonic cutter without CNN applied, the proposed method without FFT post-processing, and the proposed method with FFT post-processing. Moreover, Table~\ref{table:2} shows the comparison data of DTW. The DTW of ConvLSTM method was average 16.29 pixels with the standard deviation of 4.19 pixels, whereas the DTW of our proposed CNN method without the FFT post-processing was average 5.95 pixels with the standard deviation of 3.67 pixels, and the obtained data by the proposed CNN method with the FFT process were 1.7 pixels with 0.76 pixels deviation.

      \begin{table}[h]
         \caption{Ultrasonic cutter’s Incision Trajectory’s DTW, Values are mean ± std.}
         \label{table:2}
        \centering
          \begin{tabular}{   c    c     c     c    }
           \hline
           \textbf{Prediciton Method}& \textbf{Ave DTW [px]} \\
           \hline \hline
           Intensity Centroid &16.29±4.19 \\ 
           Proposed w/o FFT  &5.95±3.67 \\ 
             \textbf{Proposed w/ FFT} & \textbf{1.7±0.76} \\ 
             \hline
          \end{tabular}
        \end{table}

\section{Discussion}
\begin{figure}
        \centering        \includegraphics[width=0.5\textwidth]{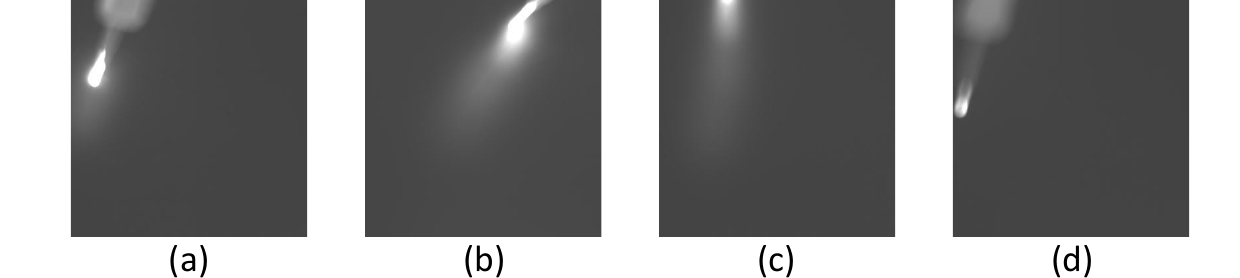}
        \caption{Examples of the false prediction images by employing electric knife: First category, the prediction was one frame earlier or later than the labeled (a)contact moment and(b) the disengagement moment. Second category, (c)The electric knife’s position at the edge of the image, which was predicted as “no contacting”. Third category, (d) The electric knife was out of focus, which was predicted as ”contacting”.}
        \label{Fig. 15}
    \end{figure}

The result of CNN prediction by employing the electric knife shows an accuracy of 97.2~\%. Although the accuracy of 97.2~\% is remarkable, a higher accuracy is expected. We thereby conducted a further analysis to interpret an issue of insufficient accuracy. The total of 18 incorrect prediction images from the results of the electric knife’s trajectories dataset were analyzed and classified into three categories. Detailed information is described as follows. In the first category, the prediction was one frame earlier or later than the contact or the disengagement moment. 11 images were predicted incorrectly, which was 61.1~\% among all of the incorrectly predicted images and 1.7~\% among all of the testing images. In the second category, the electric knife’s position was located at the edge of the image. 5 images were predicted incorrectly, which was 27.8~\% among all of the incorrectly predicted images and 0.8~\% among all testing images. In the third category, the electric knife was out of focus. 2 images were predicted incorrectly, which was 11.1~\% among all of the incorrectly predicted images and 0.3~\% among all of the testing images. On the basis of the above results, the issue may be improved by either limiting the available prediction area based on the camera view (the second category) or adjusting the camera to a wider coverage field of the view (the third category) while it can be neglected due to only a very few pixels error from the ground truth trajectory (the first category). The examples of thermal images in the above three categories are shown in Figure~\ref{Fig. 15}. 

In order to evaluate the robustness, the 5 fold cross validation was conducted to the pre-trained model of Efficientnet b4 by dividing the training, validation and test set into the ratio of 7:2:1. The results are shown in Table~\ref{table:3} with the highest accuracy 98.2~\%, the lowest accuracy 96.7~\%, and the average accuracy 97.3~\%. These data revealed the fact that all the highest, lowest, and average accuracy are close to each other, implying a trustful accuracy (97~\%) obtained in the CNN prediction by employing the electric knife.

\begin{figure}
        \centering        \includegraphics[width=0.5\textwidth]{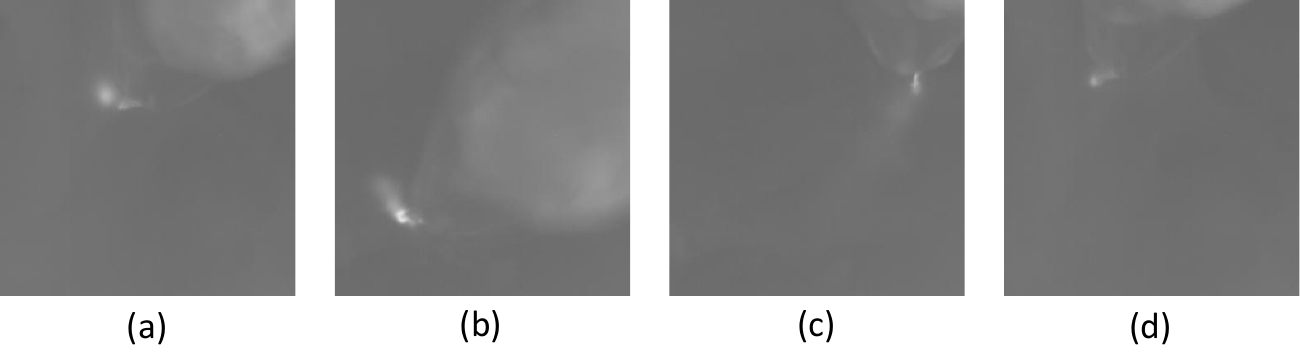}
        \caption{Examples of the false prediction images by ultrasonic cutter: First category, the prediction was one frame earlier or later than the labeled (a) contact moment and (b) the disengagement moment. Second category, (c)The angle of edge view of the ultrasonic cutter. Third category, (d) The ultrasonic cutter was a little out of focus, so the image couldn’t show a clear edge of blade.}
        \label{Fig. 16}
    \end{figure}

In terms of the results by employing the ultrasonic cutter which shows the accuracy of 93.7~\%, we also conducted a further analysis to interpret an issue of insufficient accuracy similar to that by employing the electric knife. The total of 42 incorrect prediction images from the results of the ultrasonic cutter’s trajectories dataset were analyzed and classified into three categories. Detailed information is described as follows. In the first category, the prediction was one frame earlier or later than the contact or the disengagement moment. 24 images were predicted incorrect, which was 57.1~\% among all of the incorrect predicted images and 3.6~\% among all of the testing images. In the second category, the angle of edge view of the ultrasonic cutter. 9 images were predicted incorrect, which was 21.4~\% among all of the incorrect predicted images and 1.3~\% among all of the testing images. In the third category, the ultrasonic cutter was a little out of focus, so the image could not show a clear edge of the blade. 9 images were predicted incorrect, which was 21.4~\% among all of the incorrect predicted images and 1.3~\% among all of the testing images. The examples of thermal images in the above three categories are shown in Figure~\ref{Fig. 16} on the basis of the above results, the issue may be improved by either applying multiple cameras at different angles (the second category) or adjusting the camera to a wider coverage field of the view (the third category) while it can be neglected due to only a very few pixels error from the ground truth trajectory (the first category). 

In order to evaluate the robustness, the 5 fold cross validation was conducted to the pre-trained model of Efficientnet b4 by dividing the training, validation and test into the ratio of 6:2:1. The results are shown in Table~\ref{table:4} with the highest accuracy 94.5~\%, the lowest accuracy 90.9~\% and the average accuracy 92.7~\%. These data revealed the fact that all the highest, lowest, and average accuracy are close to each other, implying a trustful accuracy (93.7~\%) obtained in the CNN prediction by employing the ultrasonic cutter.

       \begin{table}[h]
         \caption{Electric knife’s "contacting" and "no contacting" cross-validation result.}
         \label{table:3}
        \centering
          \begin{tabular}{   c    c     c     c    }
           \hline
           \textbf{Fold}& \textbf{Accuracy [\%]} \\
           \hline \hline
           Fold0 &97.2 \\ 
           Fold1 &96.7 \\ 
           Fold2 &98.2 \\ 
           Fold3 &97.5 \\ 
           Fold4 &97.1 \\ \hline
            Average  &  97.3 \\ \hline
          \end{tabular}
        \end{table}
        
       \begin{table}[h]
         \caption{Ultrasnoic cutter’s "contacting" and "no contacting" cross-validation result.}
         \label{table:4}
        \centering
          \begin{tabular}{   c    c     c     c    }
           \hline
           \textbf{Fold}& \textbf{Accuracy [\%]} \\
           \hline \hline
           Fold0 &93.7 \\ 
           Fold1 &93.6 \\ 
           Fold2 &90.9 \\ 
           Fold3 &94.5 \\ 
           Fold4 &91.0 \\ \hline
            Average  &  92.7 \\ \hline
          \end{tabular}
        \end{table}

The above analyzes suggest that most incorrect predictions only occurred in one frame earlier or later than the contact or the disengagement moment, which corresponds to the DTW results, and can suggest that the average output error of the trajectories is less than 1 pixel in the electric knife dataset and 2 pixels in the ultrasonic cutter’s dataset at 64×64 resolution, which represent roughly 0.5~mm and 1~mm of the error separately. Although the DTW result by employing the ultrasonic cutter with FFT reveals a considerably better result than that without FFT. It is worth noting that the DTW result by employing the electric knife with FFT is slightly worse than that without FFT. Even so, since both of the DTW results by employing the electric knife with and without FFT reach a remarkable low value, and the DTW difference in these two cases is within an allowable range, which makes the prediction method acceptable. According to these results, the proposed method not only has overcome the problem of missing trajectories that occur in the ConvLSTM method but also has outperformed a great improvement on accuracy. 

\section{Conclusions}
   We developed a frame extracting method by combining an approach of training CNN to classify the thermal images of the electric tools’ cutting status with another approach of acquiring the highest temperature point in each frame. We ensured the possibility of CNN distinguishing whether or not the electric tools contacting the tissue, and confirmed that the proposed method has not only overcome the problem of missing trajectories which usually occurs in the ConvLSTM method but also achieved a remarkable improvement of the accuracy. Moreover, our method requires neither the removal of the frames containing the electric knife nor the subsequent frames. Accordingly, this method can not only be applied to the higher temperature thermal diffusion by the electric tools such as electric knife, which can reach to a temperature over 80~°C but also can be applied to the lower temperature thermal diffusion by the electric tools such as the ultrasonic cutter, which can reach a temperature as low as 40~°C below. The obtainable results imply a possible practical application by ongoing the work in validating a larger number of data under various conditions, such as using a CUSA ultrasonic knife as an alternative tool by trained surgeons, as well as achieving a multi-camera system to form the trajectory up to three dimensions rather than two dimensions. In the future, the proposed method is expected to be integrated with a surgical navigation system by acquiring more clinical data.

\section*{Ethics statement}
This study did not involve experiments with human participants or live vertebrate animals. All experiments were performed using edible meat products obtained from commercial sources, and no ethical approval was required according to institutional and national guidelines.

\section*{Declaration of competing interest}

\section*{Funding}
This work was partly supported by grants from JSPS KAKENHI (JP22K08879, JP24K02969, JP24K22316, JP26K11499, JP26K15705).


\section*{CRediT authorship contribution statement}

\section*{Declaration of generative AI in scientific writing}
During the preparation of this work, the authors used deepL, Writefull, and GPT in order to proofread the manuscript because they are not native English speakers. After using this service, the authors reviewed and edited the content as needed and take full responsibility for the content of the publication.

\end{document}